\documentstyle[aps,prl,graphicx]{revtex}
\begin{document}
\draft
\twocolumn[\hsize\textwidth\columnwidth\hsize\csname 
@twocolumnfalse\endcsname
\title{Magnetic and superconducting instabilities of the Hubbard model at the van Hove filling}

\author{Carsten Honerkamp$^{1}$ and Manfred Salmhofer$^{2}$}

\address{$^{1}$ Theoretische Physik,
ETH-H\"onggerberg, CH-8093 Z\"urich, Switzerland \\
$^{2}$ Theoretische Physik, Universit\"at Leipzig, 
D-04109 Leipzig, Germany} 
\date{September 21, 2001}
\maketitle
\begin{abstract} 
We use a novel temperature--flow renormalization group technique to analyze
magnetic and superconducting instabilities in the two-dimensional $t$-$t'$
Hubbard model for particle densities close to the van Hove filling as a
function of the next-nearest neighbor hopping $t'$. In the one-loop flow at the van Hove filling, the characteristic
temperature for the flow to strong coupling is suppressed drastically 
around $t_c'\approx -0.33t$,
suggesting a quantum critical point between $d$-wave pairing at moderate
$t'>t'_c$ and ferromagnetism for $t'<t'_c$.  Upon increasing the particle density in the latter regime the leading instability occurs in the triplet pairing channel.
\end{abstract}
\vskip1pc]
\narrowtext
In recent years the two-dimensional (2D) Hubbard model has mainly served 
as an idealized model in the efforts to understand the Mott-insulating, antiferromagnetic (AF), superconducting (SC) and more exotic properties 
of the high-$T_c$ cuprates, like the pseudo-gap or stripe phases. 
While the latter phenomena are still under current debate 
and the Mott-insulating state is generically restricted to 
larger onsite repulsion $U$, the AF and $d$-wave SC 
ground states in the Hubbard model can be understood, at least qualitatively, by calculations at weak or moderate $U$. 
Here, besides diagrammatic approaches taking into account subclasses of
diagrams, like ladder summations\cite{djs} 
and the fluctuation-exchange approximation\cite{bickers},
one-loop renormalization group (RG) methods restricted to the saddle point
regions\cite{dzialoshinski,alvarez}
and more recently taking into account the whole Fermi
surface \cite{zanchi,halboth,honerkamp} have become a popular method to study the interplay between magnetism
and superconductivity in the Hubbard model. 
They indicate that for small nearest-neighbor
hopping $t'$ and close to half-filling, the weakly coupled state
is unstable with respect to AF fluctuations, while at smaller
or larger electron densities
the RG flow exhibits a Cooper instability in the $d_{x^2-y^2}$-wave
channel\cite{halboth}.  For $t' \approx -0.25t$ to  $-0.3 t$
on the hole-doped side, the change
from AF to $d$-wave instability leads over an intermediate regime,
the {\em saddle point regime} of Ref.\ \onlinecite{honerkamp}. 

Originally the Hubbard model was introduced by Hubbard,
Gutzwiller, and Kanamori with a somewhat different motivation, 
namely as a model to describe
the metallic ferromagnetism in the transition metals\cite{vollhardt,fazekas}. Since then several attempts 
have been undertaken to clarify the conditions for itinerant
ferromagnetism within the Hubbard model and related systems. 
For several classes of one-dimensional\cite{fazekas} and higher-dimensional\cite{mielke}
 models, 
spin polarized ground states can be proven to exist, and also in the limit of
infinite dimensions a number of clear results exist \cite{ulmke,vollhardt}. 
For the 2D  $t$-$t'$ Hubbard model at large $U$ Becca and Sorella\cite{becca} 
recently found numerical evidence for a Nagaoka phase close to half-filling. 
At weak to moderate $U$ however the situation is more involved.
Hartree-Fock treatments \cite{linhirsch} and more recently $T$-matrix
approximation calculations combined with Quantum Monte Carlo\cite{hlubina} and 
parquet approaches \cite{irkhin}
indicated that for larger absolute values of the next-nearest neighbor hopping 
$t'$ the ground state should be
ferromagnetic (FM) in a certain density range around the van Hove-filling where
the Fermi surface (FS) is near the saddle points of the band dispersion.    
A one-loop scaling analysis restricted to the saddle-point regions indicated similar 
tendencies \cite{alvarez}. Nonetheless,  
one-loop RG calculations\cite{zanchi,halboth,honerkamp} derived from exact RG
equations and covering the
full FS have so far not indicated any %far failed to observe indications for 
FM ground states. Moreover the competition between FM and SC ground states
remains an open problem.

In this Letter we reinvestigate possible instabilities of the 
2D $t$-$t'$ Hubbard model towards magnetic and superconducting phases
from the weak-coupling perspective. The microscopic Hamiltonian we study is 
\[ H= -t \sum_{\mathrm{n.n.} , \, s } c^\dagger_{i,s} c_{j,s} -t'
\sum_{\mathrm{n.n.n.},\, s } c^\dagger_{i,s} c_{j,s} +U \sum_i n_{i\uparrow} n_{i 
\downarrow} \, \]
with onsite repulsion $U$ and hopping amplitudes $t$ and $t'$ between
nearest neighbors (n.n.) and next-nearest neighbors
(n.n.n.) on the 2D square lattice.
We apply a novel one-loop RG scheme 
covering the full FS which we call the temperature-flow ($T$-flow) RG scheme. 
Like the approaches in
Refs. \onlinecite{zanchi,halboth,honerkamp,salmhofer} 
it is derived from an exact RG equation. However,  no
infrared cutoff is used, and instead the temperature $T$ itself 
is the flow parameter. This
new feature allows for the first time an unbiased comparison between AF and FM
tendencies, whereas RG schemes with a flowing infrared
cutoff artificially suppress particle-hole excitations with 
low wavevectors. 
A detailed discussion of this issue
and a derivation of the $T$-flow RG equations are given in Ref.\
\onlinecite{honerkampfm}. 
Using the RG formalism, we obtain a hierarchy of differential 
equations for the one-particle irreducible $n$-point vertex functions $\Gamma_T^{(n)}$
as functions of the temperature. Integration of this system gives
the $T$-flow. We pose the initial condition  
that at a higher temperature $T_0$ the
single-particle Green's function of the system is described by
$G_0 (i \omega , \vec{k} ) = [ i \omega - \epsilon (\vec{k}) ]^{-1} $
and the interaction vertex is given by a local repulsion $U$.
This is justified if $T_0$ is large enough.
We truncate the infinite system of equations by dropping all $\Gamma_T^{(n)}$ with $n>4$.
In the present treatment we also neglect selfenergy corrections and the
frequency dependence of the vertex functions. 
This restricts the scheme to a one--loop equation for the spin-rotation invariant 
four-point vertex $\Gamma_T^{(4)}$.
Starting with weak to moderate interactions, 
we follow the flow of $\Gamma_T^{(4)}$  as $T$ decreases.

$\Gamma_T^{(4)}$ is determined by a coupling function 
$V_{T} (\vec{k}_1, \vec{k}_2, $ $\vec{k}_3)\;$ \onlinecite{honerkamp,salmhofer,honerkampfm}. 
For the numerical treatment, 
we define elongated phase space patches around lines
leading from the origin of the BZ to the $(\pm \pi,\pm \pi)$-points,
and approximate $V_{T} (\vec{k}_1,\vec{k}_2,\vec{k}_3)$ 
by a constant for all wave vectors in the same patch.
We calculate the RG flow for the subset of interaction vertices 
with one wavevector on the FS representative for each patch,
with initial condition $V_{T_0} (\vec{k}_1,\vec{k}_2,\vec{k}_3)=U$.  
Most calculations were performed using 48 patches.  
 Further we calculate several static
susceptibilities, as described in Ref. \onlinecite{honerkamp}. Here we focus
on the most divergent ones, namely 
$d_{x^2-y^2}$-wave and $p$-wave pairing
susceptibilities, the AF susceptibility $\chi_s (\vec{q}= (\pi, \pi))$ 
and the FM susceptibility $\chi_s (\vec{q} \to 0)$.
This way we analyze which classes of coupling functions and
which susceptibilities become important at low $T$.
For a large parameter range, we observe a flow to strong coupling. 
The approximations mentioned above fail when the couplings get too large. 
Therefore we stop the flow when the largest coupling
exceeds a high value larger than the bandwidth, 
e.g. $V _{T, \mathrm{max}}=18t$. 
This defines a {\em characteristic temperature} $T_c$ 
of the flow to strong coupling.
Because breaking of continuous symmetries is impossible in 2D at $T>0$, 
we interpret $T_c$ as an estimate for the temperature where 
ordering occurs 
when a small additional coupling in the third spatial direction is included. 

Here we describe the results for the RG flow of the coupling function 
for the 2D Hubbard model with initial interaction
$U=3t$ and varying value for the next-nearest neighbor hopping $t'$. The
chemical potential is fixed\cite{fn1} at the van Hove (VH) value $\mu = 4t'$
so that the FS
 always contains the saddle points at $(\pm \pi,0)$ and $(0,\pm\pi)$.  

\begin{figure} 
\begin{center} 
\includegraphics[width=.36\textwidth]{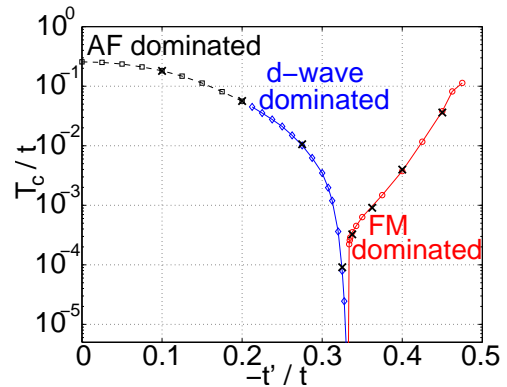}
\end{center}
\caption{
The characteristic temperature for the flow to strong coupling, 
$T_c$, versus next
nearest neighbor hopping amplitude $t'$ from a $48$-patch calculation (the crosses 
show data for 96 patches). 
The chemical potential is fixed at
the van Hove value $\mu =4t'$. $T_c$ is defined as temperature where the
couplings reach values larger than $18t$.  As 
criterion for the distinction between antiferromagnetic, $d_{x^2-y^2}$-wave
pairing and ferromagnetic regime we use
the temperature derivatives of the susceptibilities at the scale
where the couplings become larger than $10t$.}
\label{tctpvh}
\end{figure}

\begin{figure} 
\begin{center} 
\includegraphics[width=.5\textwidth]{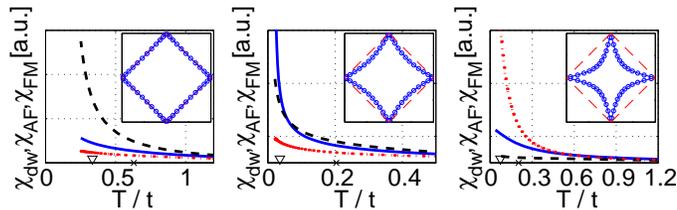}
\end{center}

\caption{Flow of the AF (dashed line), $d$-wave pairing (solid line) and FM
(dotted line) susceptibilities in  
the AF nesting regime
(left plot, $\mu=-0.01t$), $d$-wave regime (middle plot, $\mu =-t$) and FM
regime (right plot, $\mu=-1.8t$). The crosses (triangles) 
on the horizontal axes denote
the temperatures where the largest couplings reach $5t$ ($10t$).
 The insets show the Fermi surfaces and the
48 points in the Brillouin zone for which the coupling function is calculated.}
\label{sus10_180}
\end{figure}

We decrease $t'$ from $0$ to more and more negative values.
The overall behavior of the flow to strong coupling is shown in
Fig. \ref{tctpvh}. Focusing on AF, $d$-wave and FM susceptibilities, 
we observe three distinct parameter regions with typical FS shown in Fig. \ref{sus10_180}. 
The first regime is closer to half-filling, where $t'>-0.2t$. Here 
$T_c$ is relatively high and the AF susceptibility grows most strongly 
(see Fig. \ref{sus10_180}). 
The dominant scattering processes are
AF processes between FS parts connected by wavevectors $\approx (\pi , \pi)$. 
When we further decrease $t'$, $T_c$ drops continuously,
and for $t'< -0.2t$, the $d$-wave
susceptibility takes over as the leading susceptibility.  
For $t' \approx -0.25t$  and band fillings slightly larger than
the VH filling we again find a regime where the flow of $d$-wave SC and AF
processes is strongly coupled. This is the saddle point regime studied in
Ref. \onlinecite{honerkamp}. 
Here we focus on the flow when we go to even more negative 
$t'$ and at the same time adjust the chemical potential such that the FS
stays at the saddle points. Then $T_c$ drops by several orders of magnitude
 for $t' \le
-0.3t$, while for $t' \le -0.33t$ it rises again. For these values the flow to 
strong coupling is dominated by processes with small momentum transfer and the
FM susceptibility $\chi_s (\vec{q} = 0)$ is by far the most divergent
susceptibility at low temperatures (right plot in Fig. \ref{sus10_180};
note that the FM susceptibility dominates already at a higher $T$,
where the largest couplings are still well within the bandwidth.
The overall behavior strongly suggests a quantum critical
point around $t'_c \approx -0.33t$ where the ground state
changes from $d$-wave singlet SC to FM. 

Lin and Hirsch\cite{linhirsch} analyzed the
Hartree-Fock phase diagram for the $t$-$t'$ Hubbard model and found the
FM state to be stable against the AF state for $t' <  -0.324t$. 
Since they used the bare Hubbard repulsion, the $d_{x^2-y^2}$-wave phase is
absent.
The critical $t'$-value from Hartree-Fock is practically 
equivalent to the critical $t'_c
\approx -0.33t$ from the RG treatment. On the other hand in Hartree-Fock one
would find a first order transition at $t'_c$, while in the RG
 the characteristic temperature gets suppressed to lowest values, hinting at 
a quantum critical point. 
A $T$-matrix approximation (TMA) applied by Hlubina et al. \cite{hlubina}
indicated a stable fully polarized FM state beyond a critical $t' \approx
-0.43t$. The screening\cite{kanamori} in the particle-particle channel of the TMA is included in our $T$-flow scheme, which gives a somewhat 
larger window for a FM state with nonzero polarization 
at the VH filling. The density range around the VH filling where FM
dominates gets more and more narrow as $t'$ is decreased towards $t'_c$.
Recently Irkhin et al.\cite{irkhin} applied a parquet scheme with a simplified
dispersion  and found indications for 
FM at the VH filling for very similar values of $t'$.

A crucial feature of models where
large-spin ground states have been found
\cite{vollhardt,fazekas,mielke}
is a sharp peak of the density of states (DOS) {\em at the bottom} of the band.
This property is shared by the 2D $t$-$t'$ Hubbard model 
for $t' \approx -0.5t$, but at the van Hove filling, the DOS 
is also peaked at the Fermi level.
In our RG calculation the DOS {\em at the Fermi level}, 
and not that at the bottom of the band, 
is essential for the FM tendencies:
they also exist, and dominate, for smaller $|t'|$ in the range described
above, and they are strongly reduced when the Fermi level is raised above the VH
energy (see below). This sensitivity of the FM regime to the FS location 
points to another effect which is not included in the present calculation.
It is probable that the inclusion
of selfenergy effects in the flow will shift the FS and alter the
low energy DOS such that the actual FM regime might
be shifted, or changed more strongly. 
The analysis of these effects is left for future work.

Next we investigate the flow to strong coupling for densities away from the 
VH filling. The RG flow for smaller $|t'|$ is 
very similar to the flow found with the
momentum-shell techniques which has been extensively discussed in
Refs. \cite{zanchi,halboth,honerkamp}. Therefore we now focus on 
the case $t'<-0.33t$ corresponding to the FM regime at
the VH filling and search for SC tendencies when the band
filling is varied.  In the 2D Hubbard model 
the occurrence of triplet pairing in vicinity of a FM phase has
been proposed by several authors\cite{hlubina,guinea}. Our approach 
allows a more detailed analysis of this interplay.

The Cooper pair scattering $V_T(\vec{k},-\vec{k},\vec{k}')$ transforms
according to one of the 5 irreducible representations of the point group of
the 2D square lattice. Among those there is the one-dimensional representation with
$d_{x^2-y^2}$-symmetry important  for smaller $|t'|$ and the 
two-dimensional representation transforming like $p_x$ and $p_y$. In a
solution of the BCS gap equation, components belonging to different
representations do not mix, while e.g. within the $p$-representation
components $p_x \propto \cos \theta$ and $p_y \propto \sin \theta$ and
corresponding higher harmonics $\propto \cos 3 \theta$ and $\propto 
\sin 3\theta$ (also called $f$-wave) will occur together.

\begin{figure} 
\begin{center} 
\includegraphics[width=.49\textwidth]{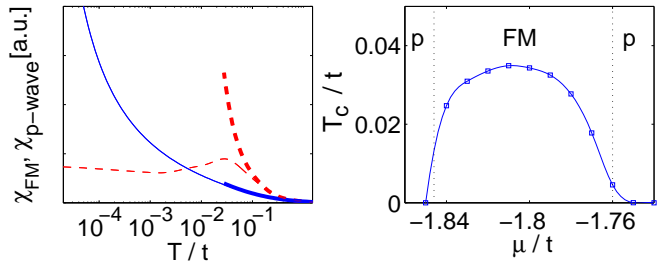}
\end{center}
\caption{Left:
Flow of ferromagnetic (solid lines) and $p_x$-wave (dashed lines) susceptibilities
 for $t'=-0.45t$ at $\mu=-1.78t$ (thick lines) and $\mu=-1.74t$ (thin
lines). Right: Characteristic
temperature $T_c$ where the largest couplings reach $18t$ versus
chemical potential $\mu$.}
\label{tp45pw}
\end{figure}

\begin{figure} 
\begin{center} 
\includegraphics[width=.5\textwidth]{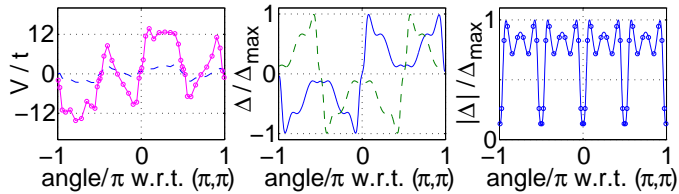}
\end{center}
\caption{Left plot:  Pair scattering
$V(\vec{k},-\vec{k},\vec{k}_3)$ with $\vec{k}$ varying around the Fermi
surface for $t'=-0.45t$ and $\mu=-1.71$ at $T=1.1\cdot 10^{-6} t$.
For the solid line $\vec{k}_3$ is fixed close to the
saddle point $(-\pi,0)$ while for the line with the dots, $\vec{k}_3 \approx
(-1.06,-0.86)$ is near the zone diagonal. 
Middle plot: Real and imaginary part of the
SC gap function obtained from the solution of the BCS gap
equation with the rescaled pair scattering from the RG flow. Right plot:
gap magnitude. }
\label{pwpscat}
\end{figure}

As shown in the right plot of
Fig. \ref{tp45pw} for $t'=-0.45t$, the critical  temperature
drops by several orders of magnitude when we increase the particle density per 
site from the VH value $\langle n \rangle \approx 0.47$ at $\mu=-1.8t$
to $\langle n \rangle \approx 0.58$ at $\mu=-1.7t$. 
Further upon moving away from the VH filling, the growth of the
FM susceptibility gets cut off and the $p$-wave triplet SC
susceptibilities with symmetry $p_x\propto \cos \theta$ or $p_y\propto \sin
\theta$ diverge at low
temperature. Higher order harmonics $\propto \cos 3 \theta$ and $\sin
3\theta$ diverge in a weaker fashion.
The pair scattering at low $T$ in
this parameter range is shown in Fig.\ref{pwpscat}. 
One nicely observes that the pair scattering
involving particles close to the saddle points is suppressed as the odd-parity 
nature of the $p$-wave pairing requires an opposite sign  of the pair
scattering $V_T (\vec{k}, -\vec{k}, \vec{k}')$ e.g. for $\vec{k}' \approx (0,
\pi)$ and $\vec{k}' \approx (0, -\pi)$.
Comparing the temperature derivatives of FM and
SC susceptibilities when the leading couplings become larger than
$12t$, the transition from the FM to the $p$-wave SC
regime occurs at a particle density $\langle n \rangle \approx 0.52$ per
site or $\mu=-1.76t$. Again, just as in the interplay between AF and $d_{x^2-y^2}$
SC fluctuations for small $|t'|$, 
our one-loop flow to strong coupling exhibits a smooth evolution 
from the FM dominated to the $p$-wave dominated instability. This
suggests a transition between the two types of
ordered states as a function of the band filling 
provided that symmetry-breaking at $T>0$ becomes 
possible in a three-dimensional environment. From our analysis the transition
appears to be first order, but additional interaction effects beyond our
one-loop calculation could change that. 

With our method we cannot calculate the SC gap function
directly, but we can use the pair scattering obtained from the RG flow 
as a pair potential in the BCS gap equation. 
In our case,  
it is plausible to assume\cite{rice} that the SC gap function will be
a nodeless superposition of the two components with
symmetries $p_x$ and $p_y$, e.g. given by
$\Delta (\vec{k}) = \Delta_0 (\vec{k}) \, (k_x + i k_y) $, 
which maximizes the condensation energy.  
The real--valued prefactor $ \Delta_0 (\vec{k})$ takes care of the 
anisotropy within symmetry-related FS parts. The direction of the Cooper pair spin
remains indeterminate in the absence of spin-orbit coupling. 
The strongly angle-dependent 
gap function calculated with
the rescaled pair scattering taken from the RG flow is shown in Fig. \ref{pwpscat}.
A similar SC order
parameter is under debate for the superconducting state of Sr$_2$RuO$_4$
\cite{physicstoday}. In this quasi-2D system one of the three
Fermi surfaces has a similar shape as the case studied above. 
It will be interesting to investigate the RG flow and the
SC properties  suggested by that
in a more realistic three-band model.

In conclusion, using the new temperature-flow RG scheme, 
we have given 
a comprehensive analysis of the flow to strong coupling for the 2D $t$-$t'$
repulsive Hubbard model on the square lattice. 
The flow to strong coupling close to half-filling for
 small next nearest neighbor hopping $t'$ with its AF and
$d_{x^2-y^2}$-pairing regimes is qualitatively similar to the RG flow 
known from the momentum-shell techniques\cite{zanchi,halboth,honerkamp}. 
For larger absolute values of $t'$ beyond a quantum critical point at
$t'_c \approx -0.33t$ 
the flow to strong coupling is dominated by FM fluctuations.
This has not been found with the momentum-shell schemes used
previously\cite{zanchi,halboth,honerkamp} for the reasons explained in
Ref. \onlinecite{honerkampfm}.
Finally, for 
larger $|t'|$, when the electron density is increased slightly above the VH
density, the FM tendencies get cut off at low T
and the flow to strong coupling is dominated by triplet
SC correlations suggesting a highly anisotropic energy gap.  

We are grateful to T.M. Rice, M. Sigrist, R. Hlubina, M.E. Zhitomirsky and
D. Vollhardt for valuable discussions. 
C.H. acknowledges financial support by the Swiss National Science Foundation.


\begin{thebibliography}{99}
\bibitem{fn1} A flow where the particle number  is fixed instead of $\mu$
gives qualitatively similar results. The same holds for smaller initial
interactions, e.g. $U=2.5t$.
\end{thebibliography}
\end{document}